\documentclass[fleqn,10pt]{wlscirep}
\usepackage[utf8]{inputenc}
\usepackage[T1]{fontenc}
\title{BrainNet: A Multi-Person Brain-to-Brain Interface for Direct Collaboration Between Brains}

\author[1]{Linxing Jiang}
\author[2,3,6,7]{Andrea Stocco}
\author[4,5]{Darby M. Losey}
\author[2,3]{Justin A. Abernethy}
\author[2,3,6,7]{Chantel S. Prat}
\author[1,6,7,]{Rajesh P. N. Rao}
\affil[1]{University of Washington, Paul G. Allen School of Computer Science \& Engineering, Seattle,
WA 98195, USA}
\affil[2]{University of Washington, Department of Psychology, Seattle, WA 98195, USA}
\affil[3]{University of Washington, Institute for Learning and Brain Sciences, Seattle, WA 98195, USA}
\affil[4]{Carnegie Mellon University, Joint Program in Neural Computation \& Machine Learning, Pittsburgh, PA 15213, USA}
\affil[5]{Carnegie Mellon University, Center for the Neural Basis of Cognition, Pittsburgh, PA 15213, USA}
\affil[6]{University of Washington Institute for Neuroengineering, Seattle, WA 98195, USA}
\affil[7]{University of Washington, Center for Neurotechnology, Seattle, WA 98195, USA}

\keywords{Brain-Computer Interface (BCI), Brain-to-Brain Interface}

\begin{abstract}

We present BrainNet which, to our knowledge, is the first multi-person non-invasive direct brain-to-brain interface for collaborative problem solving. The interface combines electroencephalography (EEG) to record brain signals and transcranial magnetic stimulation (TMS) to deliver information noninvasively to the brain. The interface allows three human subjects to collaborate and solve a task using direct brain-to-brain communication. Two of the three subjects are designated as “Senders” whose brain signals are decoded using real-time EEG data analysis. The decoding process extracts each Sender’s decision about whether to rotate a block in a Tetris-like game before it is dropped to fill a line. The Senders’ decisions are transmitted via the Internet to the brain of a third subject, the “Receiver,” who cannot see the game screen. The Senders' decisions are delivered to the Receiver’s brain via magnetic stimulation of the occipital cortex. The Receiver integrates the information received from the two Senders and uses an EEG interface to make a decision about either turning the block or keeping it in the same orientation. A second round of the game provides an additional chance for the Senders to evaluate the Receiver’s decision and send feedback to the Receiver’s brain, and for the Receiver to rectify a possible incorrect decision made in the first round. We evaluated the performance of BrainNet in terms of (1) Group-level performance during the game, (2) True/False positive rates of subjects’ decisions, and (3) Mutual information between subjects. Five groups, each with three human subjects, successfully used BrainNet to perform the collaborative task, with an average accuracy of 81.25\%. Furthermore, by varying the information reliability of the Senders by artificially injecting noise into one Sender’s signal, we investigated how the Receiver learns to integrate noisy signals in order to make a correct decision. We found that like conventional social networks, BrainNet allows Receivers to learn to trust the Sender who is more reliable, in this case, based solely on the information transmitted directly to their brains. Our results point the way to future brain-to-brain interfaces that enable cooperative problem solving by humans using a “social network” of connected brains.

\end{abstract}
\begin{document}

\flushbottom
\maketitle

\thispagestyle{empty}

\section*{Introduction}

Direct brain-to-brain interfaces (BBIs) in humans\cite{Rao:2014,RaoPilot2013, Min:2010,Grau:2014,Stocco:2015} are interfaces which combine neuroimaging and neurostimulation methods to extract and deliver information between brains, allowing direct brain-to-brain communication. A BBI extracts specific content from the neural signals of a “Sender” brain, digitizes it, and delivers it to a “Receiver” brain. Because of ethical and safety considerations, existing human BBIs rely on non-invasive technologies, typically electroencephalography (EEG), to record neural activity and transcranial magnetic stimulation (TMS) to deliver information to the brain. For example, the first human BBI demonstrated by Rao and colleagues in 2013\cite{RaoPilot2013} decoded motor intention signals using EEG in the Sender and conveyed the intention via TMS directly to the motor cortex of the Receiver to complete a visual-motor task \cite{Rao:2014}. Stocco and colleagues\cite{Stocco:2015} extended these results by showing that a Sender and a Receiver can iteratively exchange information using a BBI to identify an unknown object from a list, using a question-and-answer paradigm akin to “20 Questions.” Grau and colleagues\cite{Grau:2014} proposed a related but offline non-iterative BBI. 

Early interest in human BBIs came from the potential for expanding human communication and social interaction capabilities \cite{RaoStocco:2014,Dingemanse:2017,Kyriazis:2015,Hongladarom2015,Montague2002}. However, previous BBIs have lacked several key features of real-world human communication. First, the degree of interactivity has been minimal; for example, in the case of the ``20 Questions'' BBI \cite{Stocco:2015}, the Sender only responds to the question the Receiver chooses, and the Receiver’s performance does not affect the Sender’s decision. Second, their interface required \emph{physical} action: the Receiver touched the screen to select a question. Thus, the communication loop was completed via a motor output channel rather than a brain interface. Third, all past human BBIs have only allowed two subjects. Human communication, on the other hand, has become increasingly dominated by means such as social media that allow multiple parties to interact in a network. The potential for BBIs allowing interactions between multiple humans has previously been theorized\cite{Min:2010,Nicolelis:2011} but not demonstrated. 

Here, we present BrainNet (Figure~\ref{fig:brainnet}), a next-generation BBI that addresses many of the limitations of past BBIs. First, BrainNet is designed to be a BBI for more than two human subjects; its current implementation allows two Senders and one Receiver to communicate, but it can be readily scaled up to large numbers of Senders. The Senders have the same role in observing the current state of the task and conveying their decisions to the Receiver. The Receiver has the role of integrating these independent decisions and deciding on a course of action. Second, BrainNet's design incorporates a second round of interactions between the Senders and the Receiver, so that the action of the Receiver in the first round can be perceived by the Senders, giving them a second chance to convey (potentially corrective) decisions to the Receiver. Third, the Receiver is equipped with both TMS (to receive Senders’ decisions) and EEG (to perform an action in the task), thereby completely eliminating the need to use any physical movements to convey information. We report results from five groups, each with three human subjects (henceforth, ``triad''), who successfully used BrainNet to perform a collaborative task based on a Tetris-like game.

An important feature of communication in social networks is deciding which sources of information to pay attention to when deciding on a course of action\cite{Bakshy:2012}. To investigate whether BrainNet allows such a capability, we additionally explored whether the Receiver can learn the reliability of each Sender over the course of their brain-to-brain interactions. We varied the reliability of the signal from one Sender compared to the other by injecting noise into the signals from one randomly chosen Sender. Our results show that like conventional social networks, BrainNet allows a Receiver to learn to trust the Sender who is more reliable, i.e., whose signal quality is not affected by our manipulation.  

\section*{Results}

To measure the direct brain-to-brain communication capabilities of BrainNet, we asked each triad of participants to perform 16 trials of an iterative Tetris-like game. In each trial, one participant, designated as the Receiver, is in charge of deciding whether or not to rotate a block before it drops to fill a gap in a line at the bottom of the screen. Critically, the Receiver is prevented from seeing the bottom part of the screen and must rely on the counsel of the other two participants, designated as the Senders, who can see the screen in its entirety. These Senders are tasked with making the correct decision (rotate or not) based on the shape of the current block and the gap at the bottom, and informing the Receiver of the decision via the brain-to-brain interface. All members of the triad communicate their decisions through an EEG-based interface using steady state visually evoked potentials (SSVEPs; see Methods). The Senders' decisions are delivered to the Receiver through two TMS pulses delivered sequentially to the occipital cortex, eliciting a phosphene for a "yes" decision or no phosphene for a "no" rotation decision for each Sender (see Methods). Each trial is composed of two rounds: the first round is as described above; after the first round, the Senders are given the opportunity to examine the Receiver's decision, shown on their screen as the block (now potentially rotated) mid-way through its fall. The Senders are then given another chance to make new (possibly corrective) suggestions to the Receiver through the brain-to-brain interface.  A successful completion of a trial thus requires accurate communication between the Senders and the Receiver across these two rounds (see Figure~\ref{fig:two-step}). Further, to examine the issue of reliability of the Senders, our software randomly chooses one Sender to be less reliable by making the decision sent to the Receiver from that Sender incorrect in ten out of sixteen trials. The order of trials requiring the block to be rotated and trials not requiring rotation was pseudo-randomized, with the constraint that each half of the session contained 4 rotation and 4 non-rotation trials. Trials 8-12 for the first triad were excluded from all analysis due to a problem with the timestamp routine. We analyzed both the EEG and behavioral data from the subjects in the remaining trials.

\subsection*{Overall Performance}

\begin{figure}[ht]
\centering
\includegraphics[width=\linewidth]{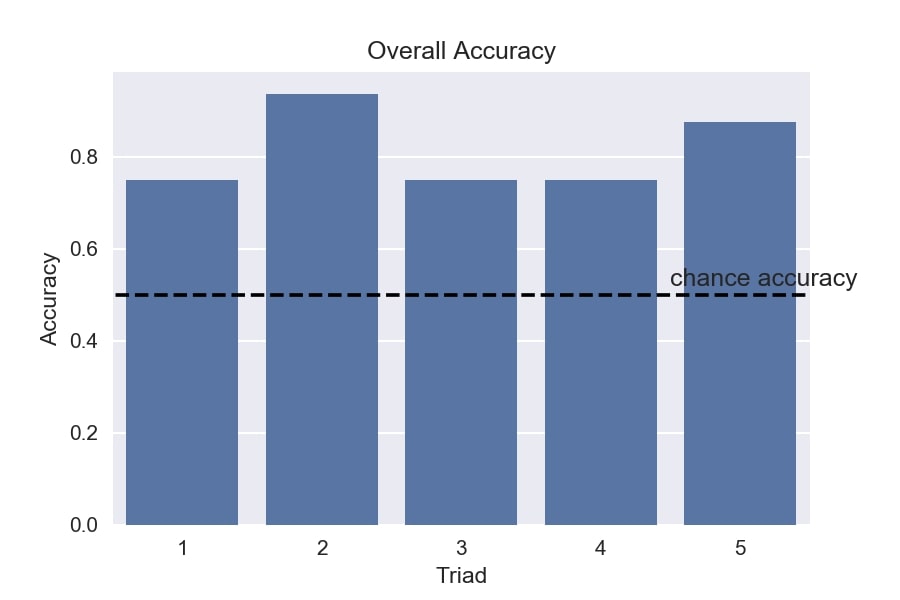}
\caption{\textbf{Performance of Triads of Participants using BrainNet.} The plot shows accuracy achieved by each of the five triads of participants. Accuracy was defined as the proportion of correct block rotations achieved by the triad. The dashed line shows the theoretical chance accuracy (0.5).}
\label{fig:accuracy}
\end{figure}

The simplest measure of overall performance of the interface is the proportion of correct block rotations (equivalently, the proportion of number of lines cleared, or the proportion of the maximum theoretical total score, i.e. 16 points, achieved) for each of the five triads of participants. Figure~\ref{fig:accuracy} shows the results. The mean accuracy across all triads was 0.8125, corresponding to 13 correct trials out of 16. A corresponding $p$-value was calculated using the binomial distribution, which confirmed that the mean performance was indeed higher than expected by chance ($p$ = .002).

\begin{figure}[ht]
\centering
\includegraphics[height=.8\textheight]{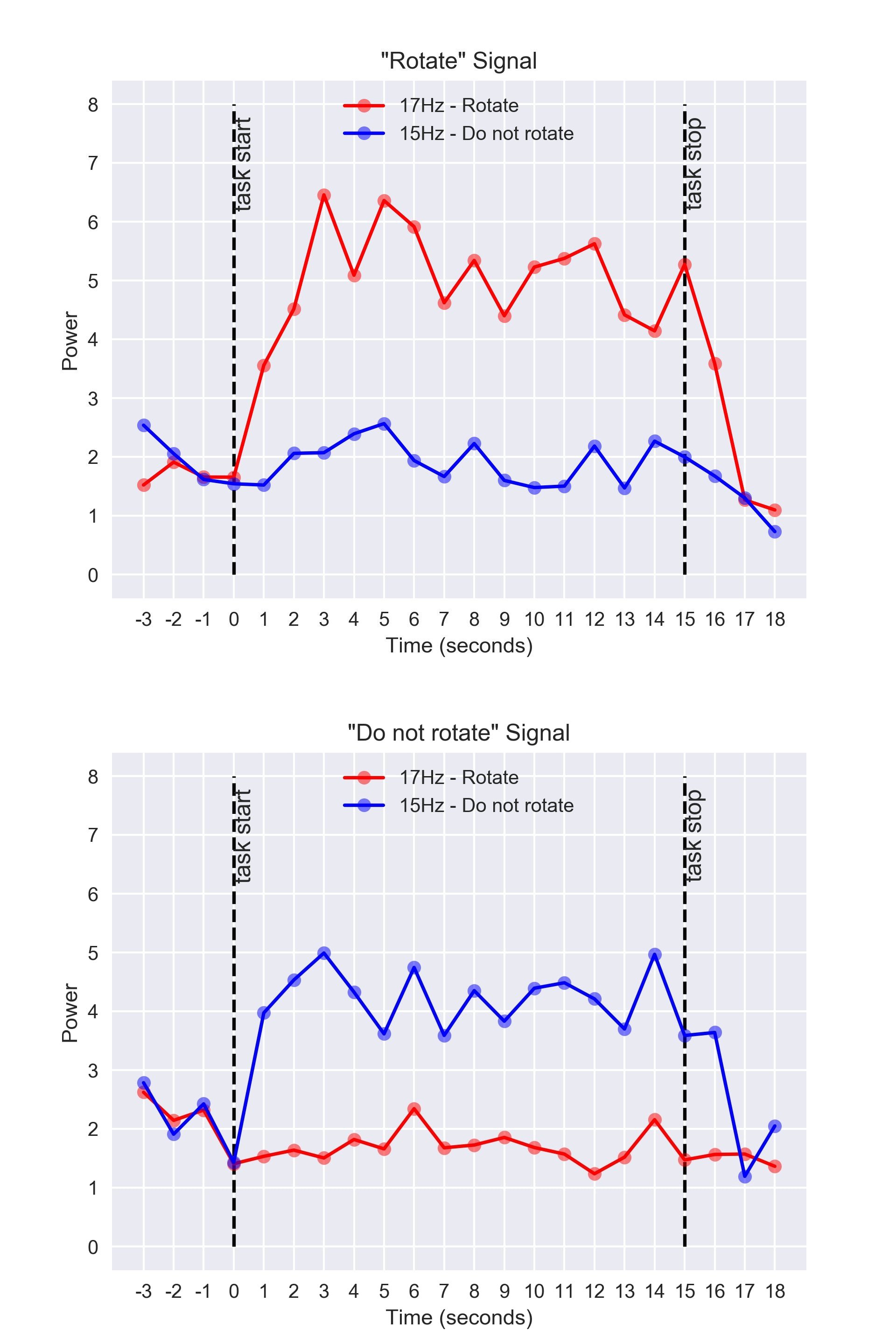}
\caption{\textbf{Average Power Spectra of EEG Signals across Subjects during the SSVEP Task.} Power values were averaged across one-second epochs and across subjects. The plots show the average power values during the SSVEP task (between dashed lines) and for comparison, the power values three seconds before and after the task. Note that before and after the task, the power values overlap for the two frequencies, whereas during the task, the power of the frequency corresponding to the correct answer is significantly larger.}
\label{fig:ssvep}
\end{figure}

Another important metric is the mean performance of participants in the SSVEP task since both Senders and the Receiver in each triad had to use this method to share information. In the task, subjects focused their attention on a 17 Hz flashing LED to indicate a ``Rotate'' decision and a 15 Hz flashing LED to indicate a ``Do Not Rotate'' decision. Figure~\ref{fig:ssvep} shows that before and after the SSVEP task, the 17 Hz and 15 Hz average power values overlap, whereas during the task, the average power of the frequency corresponding to the correct answer in the trial is significantly larger than that of the frequency corresponding to the wrong answer (two-sample $t$-test; $t$(15) = 9.709, $p$ < 0.0001 for "Rotate" signal; $t(15)$ = 10.725, $p$ < 0.0001 for "Do Not Rotate" signal).
Since our SSVEP classifier compares the magnitude of power values to decode a Sender's decision, the large difference in power values implies good performance for our EEG-based brain interfaces. 

\begin{figure}[ht]
\centering
\includegraphics[width=\linewidth]{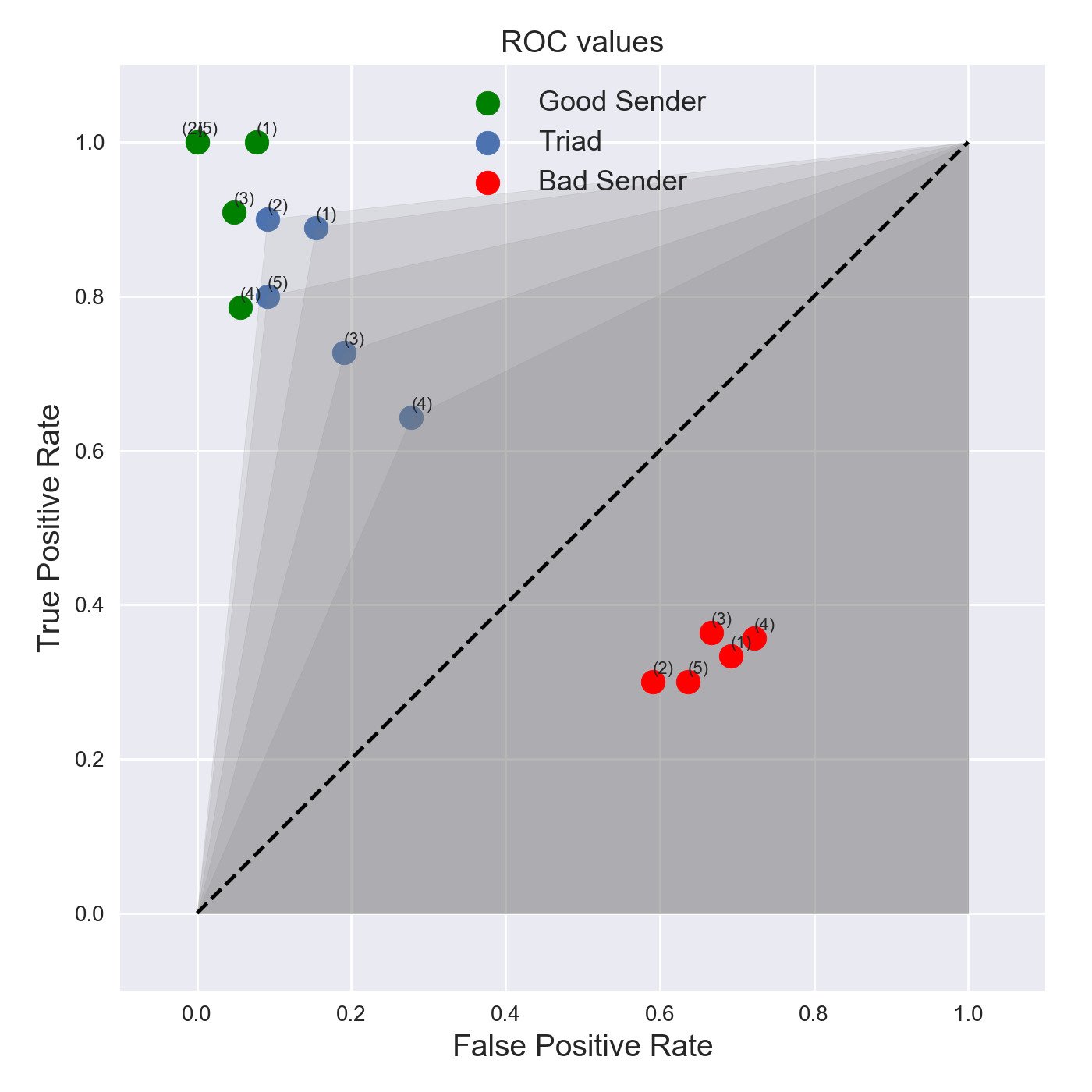}
\caption{\textbf{ROC Curves for the Five Triads of Participants.} The plot shows the overall performance of each triad (blue dots) as well as the performances of the two types of Senders ("Good" versus "Bad") in each triad (green and red dots). See text for details on the experimental design used to create a ``Good'' versus ``Bad'' Sender. The superscript on each dot denotes the triad number. Shaded areas represent the area under the curve (AUC) for each triad's ROC curve. The dashed line denotes chance performance.}
\label{fig:roc}
\end{figure}

As noted in previous studies from our group \cite{Rao:2014,Stocco:2015,Losey:2016}, raw accuracy can be an inadequate measure of performance because it does not differentiate the kind of mistakes being made, i.e., whether they are misses or false positives. A better measure of performance can be obtained by calculating each triad’s Receiver Operating Characteristic (ROC) curve\cite{Fawcett:2006}, which plots the True Positive Rate versus the False Positive rate, and  calculating the area under this curve (AUC; Figure~\ref{fig:roc}).  Responding uniformly at random yields an AUC of 0.5 while the AUC of an ideal observer is 1.0. Because the distribution of AUC values  is constrained between 0 and 1, it does not conform to the normal distribution. Thus, to properly conduct statistical tests of these values, we followed two complementary approaches. First, we conducted $t$-tests on the angular transformation (i.e., the arcsine square root transformation) of the AUC values, a common technique used to normalize data distributions \cite{rao1998statistical}. Second, we entered the raw, untransformed values in a Wilcoxon test, a non-parametric test with a continuity correction. Both tests confirmed that the mean AUC value of 0.83 across all triads of participants was significantly higher than the performance expected by chance (one-sample $t$-test on angular transformed data: $t$(4) = 11.366, $p$ < .001; one-sample Wilcoxon test: $V$ = 15, $p$ = 0.031).

As Figure~\ref{fig:roc} shows, the overall AUC value for each triad of brains is affected by the bad Sender's performance, but not by much. The overall AUC values are smaller than the AUC values of the good Senders (two-sample $t$-test on angular transformed data: $t$(4) = -2.897, $p$ = .021; Wilcoxon test: $W$ = 2, $p$ = .036) but significantly larger than those of the bad Senders (two-sample $t$-test on angular transformed data: $t$(4) = 9.184, $p$ < 0.001; Wilcoxon test: $W$ = 25, $p$ = 0.008). 

\subsection*{Mutual Information Between Participants}

An important measure of a brain-to-brain interface is the mutual information (MI)\cite{Cover:2006} transmitted between subjects, which is defined as:

$$MI(R, S) = \sum_{r\in \{0, 1\}}\sum_{s\in \{0, 1\}}p_{R, S}(r, s)\log\frac{p_{R, S}(r, s)}{p_{R}(r)p_{S}(s)}$$

\noindent where $r$ represents a decision made by the Receiver (0 or 1 corresponding to "do not rotate" or "rotate"), $s$ represents a decision made by one of the Senders, $p_{R}(r)$ represents the probability of the Receiver making the decision $r$, $p_{S}(s)$ represents the probability of one of the Senders making the decision $s$, and $p_{R, S}(r, s)$ represents the {\em joint} probability of the Receiver making the decision $r$ and a Sender making the decision $s$. Note that, in this case, chance performance corresponds to $\textrm{MI} = 0.0$ while perfect communication corresponds to $\textrm{MI} = 1.0$. Because mutual information values are also constrained between 0 and 1 and, therefore, are not normally distributed, we analyzed them using the statistical methods we applied to the AUC values (i.e., $t$-test on angular transformed data and Wilcoxon test with continuity correction).

\begin{figure}[ht]
\centering
\includegraphics[width=\linewidth]{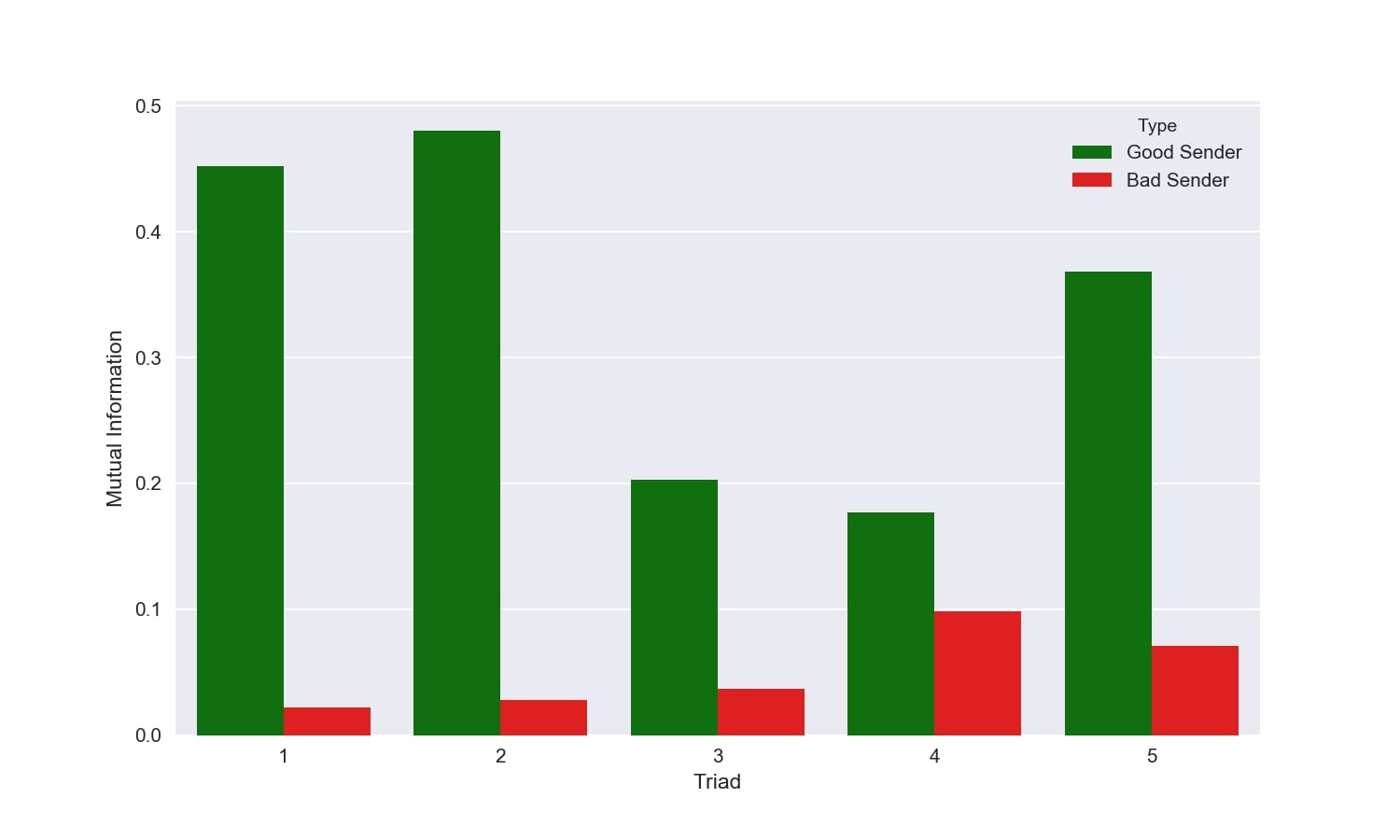}
\caption{\textbf{Mutual Information transmitted between the Senders and the Receiver.} Across all five triads of BrainNet participants, the mutual information transmitted between the Receiver and the "Good" Sender is significantly higher than that between the Receiver and the "Bad" Sender. }
\label{fig:mi}
\end{figure}

Due to our experimental design, we expect significantly higher MI values (i.e., larger amounts of information being transferred) between a good Sender and the Receiver than between a bad Sender and the Receiver. This is corroborated by our results (Figure~\ref{fig:mi}).

The information transmitted was significantly greater than the MI for chance performance for both the good Senders ($MI$ = 0.336, $t$-test on angular transformed data:  $t$(4) = 5.374, $p$ =.006; Wilcoxon test: $V$ = 15, $p$ = 0.031) and the bad Senders ($MI$ = $0.051$; $t$-test on angular transformed data:  $t$(4) = 3.544, $p$ =.024; Wilcoxon test: $V$ = 15, $p$ = 0.031). The difference between good and bad Senders was also statistically significant (two-sided $t$-test on angular transformed data, $t$(8) = 5.187, $p$ = .002; Wilcoxon test: $W$ = 0, $p$ = 0.031), with the good Senders transmitting, on average, more information than the bad Senders.

For consistency with previous studies \cite{RaoStocco:2014,Stocco:2015,Rao:2014}, we have reported {\it uncorrected} estimates of MI. Given the relatively small number of samples, uncorrected MI values might overestimate the true amount of information shared by two participants. For this reason, we used a recently proposed method \cite{Panzeri:2017} to calculate the amount of bias in our estimates. Under the conditions of our experiment, the bias $b$ can be approximated as $b = -N_R/[2 \times N_S \times \log(2)]$, with $N_R$ being the number of possible responses (in our case, $N_R=2)$ and $N_S$ the number of samples (in our case, $N_S = 32$ for each pair of participants). The bias thus estimated was found to be negligible ($b = -0.045$) and does not affect the results of any of our statistical tests.

\subsection*{Learning of Sender Reliability by Receiver}

The differences in accuracy and mutual information between the ``good'' and ``bad'' Senders in the previous section suggest that the Receiver successfully learned which of the two Senders is a more reliable source of information. Confirming that this is indeed the case would bring BrainNet a step closer to conventional social networks where users utilize differential weighting for different sources of information. To further investigate this issue, we divided each experimental session into four consecutive blocks of four trials each. We quantified the time course of the Receiver's learning process using two measures: (1) block-by-block estimates of the linear regression weights for the Receiver's decisions versus each Sender's decisions; and (2) the block-by-block correlation of decisions made by the Receiver and by each Sender \cite{Rao:2014}. Because of the small number of trials ($N=4$) in each block, the decision vectors for Senders and Receivers were created by concatenating the decisions of participants with the same role (Receiver, good Sender, or bad Sender) across the five triads; this procedure captures group-level behavior and is less sensitive to outliers. Thus,  if $R\in\mathbb{R}^{20\times 1}$ is a decision vector for the five Receivers in a four-trial block (each decision is encoded as a 0 or 1), and $S\in\mathbb{R}^{20 \times 1}$ is a decision vector for one type of Sender ("good" or "bad"), the linear regression weights $ \beta$ can be estimated using the standard pseudoinverse method\cite{Kenney:1947} as: $ \beta = (S^TS)^{-1}S^TR$. The same concatenated vectors $R$ and $S$ were also used to estimate the Pearson correlation coefficients for each four-trial block.

\begin{figure}[ht]
\centering
\includegraphics[scale=0.2]{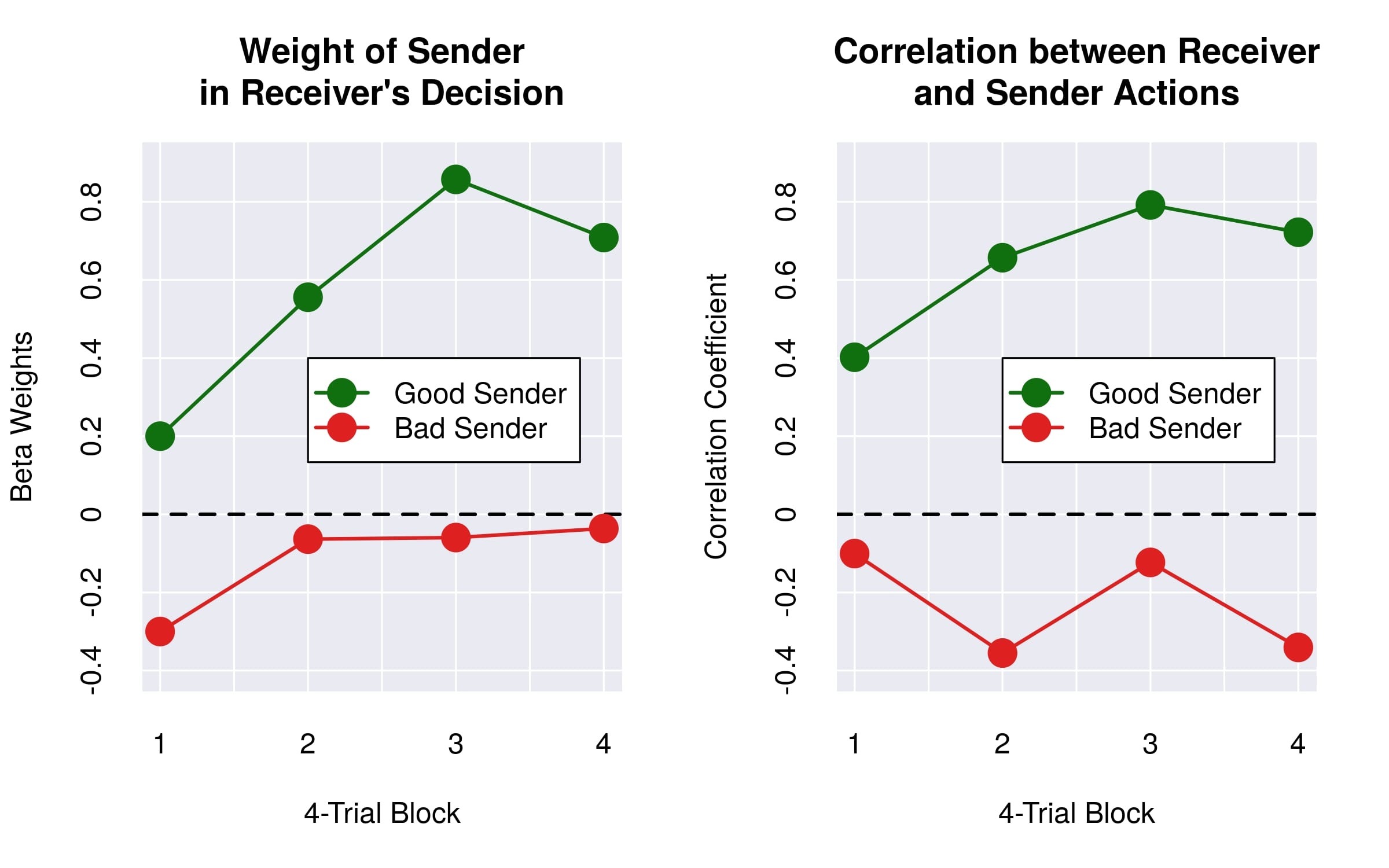}
\caption{\textbf{Quantification of Learning of Sender's Reliability by Receiver.} (Left Panel) Evolution over time of linear regression weights (Beta) for the Receivers' decision vector and decision vector for each type of Sender for each 4-trial block (see text for details). (Right Panel) Evolution over time of Pearson Correlation Coefficient between the decisions of Receivers and Senders of each type. Both plots exhibit ascending trends for the "Good" Sender but not the "Bad" Sender, suggesting that Receivers learned which Sender was more reliable during the course of their brain-to-brain interactions with the two Senders.}
\label{fig:learning}
\end{figure}

As shown in Figure~\ref{fig:learning}, the time course of both the beta weights and correlation coefficients show a steep ascending trend for the good Sender, but not for the bad Sender. To test the difference between these trends, we estimated two simple linear trend models of the relationship between each measure and the block number, one for the good Sender and one for the bad Sender. The difference between the linear trend model's slope coefficients $\beta_g$ and $\beta_b$ for the good and bad Senders respectively was then tested for statistical significance using the formula derived by Paternoster and colleagues \cite{Paternoster:1998}:

$$ Z = \frac{\beta_g - \beta_b}{\sqrt{SE\beta_g^2 + SE\beta_b^2 }} $$

where $SE\beta_g^2$ and $SE\beta_b^2$ are the variances of $\beta_g$ and $\beta_b$, respectively. The difference in linear trends was statistically significant for both measures (beta weight measure: Z = 5.87, $p$ < 0.001; correlation coefficient measure: Z = 7.31, $p$ < 0.001). These results strongly suggest that Receivers were able to learn which Sender was more reliable during the course of their brain-to-brain interactions with the two Senders.

\section*{Discussion}

This paper presents, to our knowledge, the first successful demonstration of multi-person non-invasive direct brain-to-brain interactions for collaboratively solving a task. We believe our brain-to-brain interface, which we call BrainNet, improves upon previous human brain-to-brain interfaces (BBIs) on three fronts: (1) BrainNet expands the scale of BBIs to multiple human subjects working collaboratively to solve a task. (2) BrainNet is the first BBI to combine brain recording (EEG) and brain stimulation (TMS) in a single human subject, eliminating the need to use any physical movements to convey information (although we did not explicitly instruct subjects to avoid eye movements when using the SSVEP interface, other researchers have shown that an SSVEP BCI can be operated without eye movements \cite{Kelly2005,Allison2008}). With sufficient hardware, our system can be scaled to the case where every subject can both send and receive information using the brain interface. (3) Using only the information delivered by BrainNet, Receivers are able to learn the reliability of information conveyed to their brains by other subjects and choose the more reliable sender. This makes the information exchange mediated by BrainNet similar to real-life social communication, bringing us a step closer to a ``social network of brains.''

Our results on combining information from multiple users builds on previous work in the field of brain-computer interfaces (BCIs) linking the individual contributions of more than two brains to control a computer. In humans, researchers have studied “collaborative BCIs” (rather than BBIs) that pool information from multiple human brains to improve performance in a delayed saccade-or-reach task\cite{Wang:2011}; however, subjects performed the task on different days and no brain stimulation was used to convey information directly to subjects' brains. A different study\cite{Ramakrishnan2015} demonstrated that three non-human primates can jointly control a 3D virtual avatar arm using brain signals recorded with invasive electrodes implanted in the motor cortex; again, the goal was distributing a single task across multiple individuals linked to a common BCI without encoding any neural information for feedback and interaction via stimulation. More closely related to our study is the work of Pais-Vieira {\em et al.}\cite{Pais-Vieira:2015}, who used implanted electrodes to both decode information from and transmit information to the somatosensory cortices of multiple rodents to demonstrate the possibility of distributing computations across multiple brains. The brains of the rats were linked to solve several computational problems including a weather forecasting task based on weather data from a local airport. However, the animals were entirely unaware of both the actual task being solved and of their collaboration with others; by contrast, in BrainNet, the participants are completely aware of the task and are conscious of being collaborators within a ``network of brains''. 

BrainNet could be improved in several ways: (1) From the first human BBI\cite{Rao:2014} to BrainNet, the level of information complexity has remained binary, i.e., only a bit of information is transmitted during each iteration of communication. Additionally, this low bit rate required a disproportionate amount of technical hardware and setup. To address the limitation of low bit rate, we are currently exploring the use of functional Magnetic Resonance Imaging (fMRI) to increase the bandwidth of human BBIs. Other approaches worth exploring include combining EEG and fMRI to achieve both high spatial and temporal resolution\cite{Huster:2012} for decoding, and using TMS to stimulate higher-order cortical areas to deliver more complex information such as semantic concepts. (2) We purposefully introduced a “bad” sender in BrainNet design to study whether the Receiver can learn which Sender is more reliable. It would be interesting to investigate whether the Receiver can learn the reliability of Senders in more natural scenarios where the unreliability originates from the noisy nature of a Sender's brain recordings or from a Sender's lack of knowledge, diminished attention, or even malicious intent. (3) From an implementation standpoint, BrainNet uses a typical server-client TCP protocol to transmit information between computers. However, the server is solely designed for BrainNet’s experimental task and is not a general purpose server. A cloud-based BBI server could direct information transmission between any set of devices on the BBI network and make it globally operable through the Internet, thereby allowing cloud-based interactions between brains on a global scale. Such BBIs, when developed within an ethically-grounded framework, have the potential to not only open new frontiers in human communication and collaboration but also provide a new scientific tool to explore questions in neuroscience and gain a deeper understanding of the human brain. 

\section*{Methods}

\subsection*{Participants}

Fifteen healthy participants (aged 18–35 yrs, average 22.7 yrs, eight female), took part in a controlled laboratory experiment. All participants were recruited through word of mouth, were fully informed about the experimental procedure and its potential risks and benefits, and gave written consent prior to the beginning of the experiment according to the guidelines put forth by the University of Washington. Both the experimental and the recruitment procedures were reviewed and approved by the Institutional Review Board of the University of Washington (IRB Application \#52392). The participants were divided into five groups, with each group being a triad of one participant playing the role of the “Receiver” and two playing the roles of “Senders.” To maintain their decision to participate free of any external influence, all participants received monetary compensation that was independent of their role and proportional to the total amount of time devoted to the study.

\subsection*{Experimental Task}

\begin{figure}[ht]
\centering
\includegraphics[width=\linewidth]{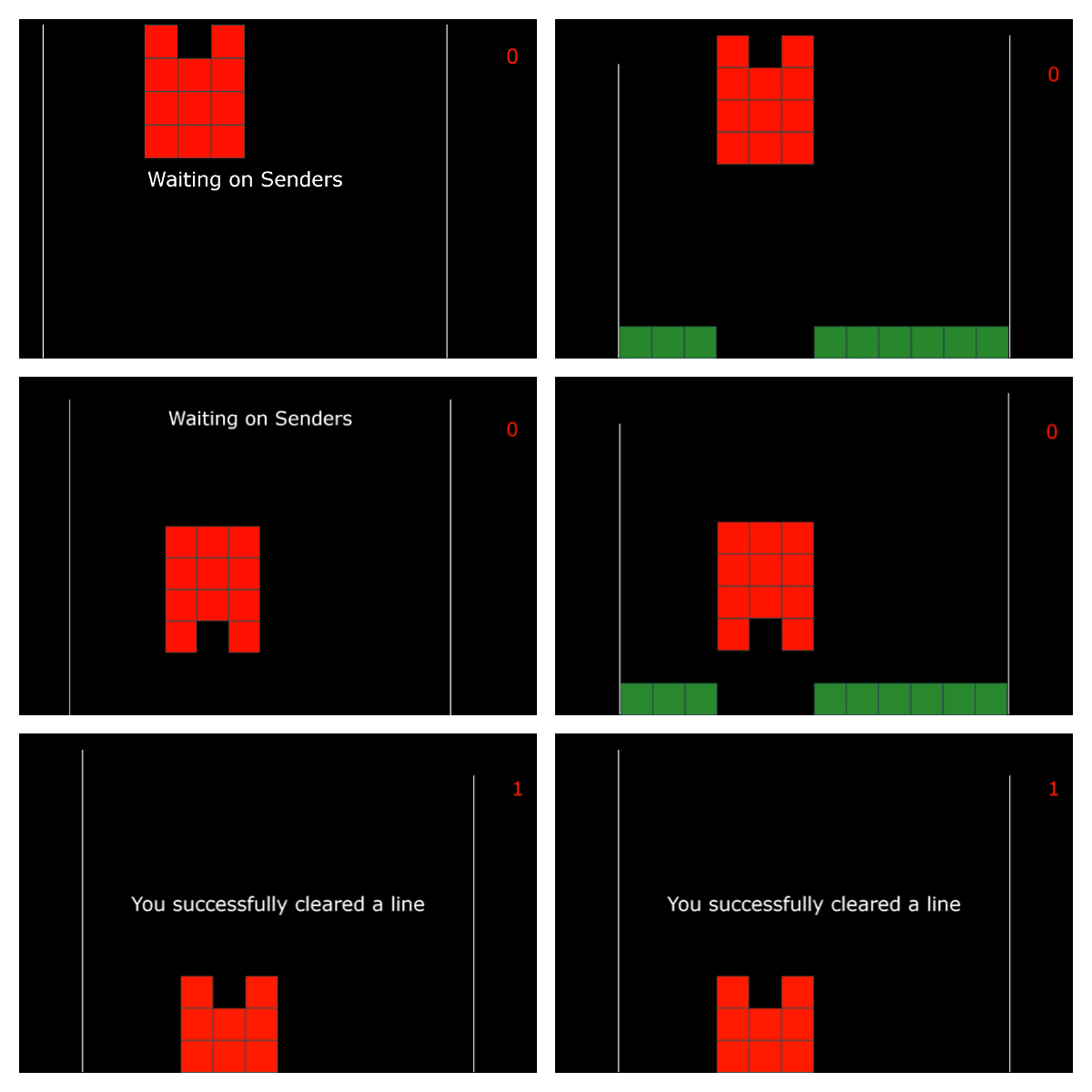}
\caption{\textbf{Examples of Screens seen by the Receiver and the Senders across Two Rounds.} The Receiver sees the three example screens on the left side and the Senders see the screens on the right side. (Top Row) Screens at the beginning of the trial. Note that the Receiver does not see the bottom line with the gap but the Senders do. The Receiver must rely on the Senders to decide whether or not the red block must be rotated to fill the gap and clear the line. (Middle Row) After the Receiver makes a decision in the first round (in this case, ``Rotate''), the game state is updated to show the rotated block. (Bottom Row) After the second round, all participants see the results of the Receiver's action and whether the line was cleared. In this example, the Receiver executed a corrective action to rotate the block again, thereby filling the gap with the bottom part of the block and clearing the line.}
\label{fig:two-step}
\end{figure}

During each session, a triad of three participants collaborated to play a simplified Tetris-like game. The game consisted of independent trials, each of which involved deciding whether or not to rotate a single block of particular shape by 180 degrees. At the bottom of the screen, there was a partially filled line whose gaps could be filled by either the top or bottom part of the block at the top of screen. The goal of the game was to achieve the highest possible score by making the correct decision to rotate or not rotate the current block so that when dropped at the end of the trial, it would fill the missing parts of the line at the bottom. We designed the task such that the actual player of the game, namely the Receiver, could only see the block at the top of screen and not the bottom line. The other two subjects, namely the Senders, could see both the block at top and the line at bottom (see Figure~\ref{fig:two-step}). Thus, the only way for the Receiver to achieve a high score was by integrating the decisions transmitted by both Senders and make his/her own decision for the game.

Each session was made of sixteen independent trials; in half of them the falling block had to be rotated and in the other half, it had to be left in the original orientation.  The order of rotation and non-rotation trials was randomized, with the constraint that each half of the session had to contain 4 rotation and 4 non-rotation trials.

Each trial comprised of two rounds of interactions between the Senders and the Receiver. Each round offered a chance to rotate the block. After the first round, the block was rotated or remained in the same orientation based on Receiver’s decision. The block then dropped halfway and the screens shown to all three subjects were updated to show the (possibly rotated) block at the halfway location (see Figure~\ref{fig:two-step}). Note that one decision is sufficient to complete the task of filling the bottom line but because of our two-step design, the Senders receive feedback on the Receiver's action in the first round and can send the Receiver new suggestions, allowing the Receiver to potentially correct a mistake made in the first round and still successfully complete a trial.

The three participants in a triad were located in different rooms in the same building on the University of Washington campus and could only communicate with each other through the brain-to-brain interface.

\subsection*{BrainNet: Multi-Person Brain-to-Brain Interface}

\begin{figure}[h]
\centering
\includegraphics[width=\linewidth]{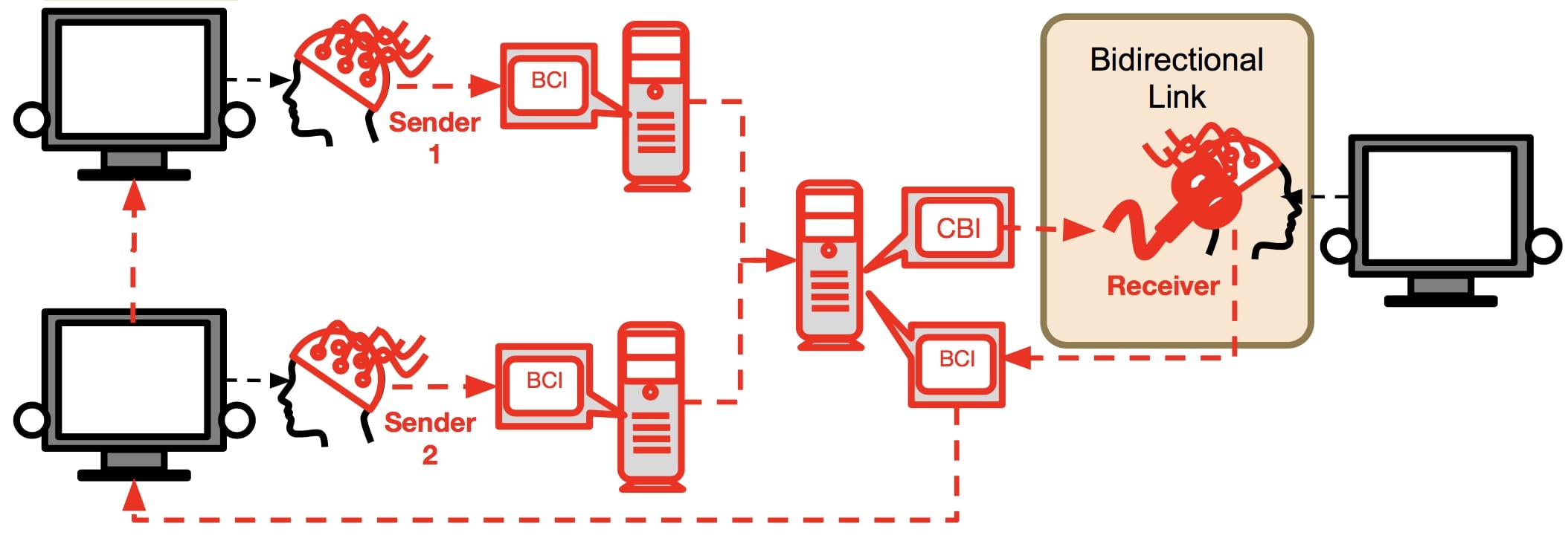}
\caption{{\bf Architecture of BrainNet.} Two participants ("Sender 1" and "Sender 2") each use a Brain-Computer Interface (BCI) based on EEG to convey information about a collaborative task (here, a Tetris-like game) directly to the brain of the third participant ("Receiver"). Information from each Sender is transmitted over the internet to the Receiver's brain via a Computer-Brain Interface (CBI) based on TMS. After consciously processing the two inputs from the Senders, the Receiver uses a BCI based on EEG to execute an action in the task. The Senders see the result of this action on their screens (the same updated game state is shown on both screens, as indicated by the red arrow from one Sender's screen to the other). The Senders then have another opportunity to convey to the Receiver's brain new information to potentially rectify an incorrect choice in the first round. While our experiment only used two rounds, BrainNet allows an arbitrary number of interactions between the Senders and the Receiver as they collaborate to solve a task. BrainNet differs from a previous interface called "Brainet"\cite{Ramakrishnan2015} which combines recordings from multiple monkey brains to perform a common motor task but is unidirectional and does not use stimulation to communicate information back to any of the brains.}
\label{fig:brainnet}
\end{figure}

\begin{figure}[ht]
\centering
\includegraphics[width=\linewidth]{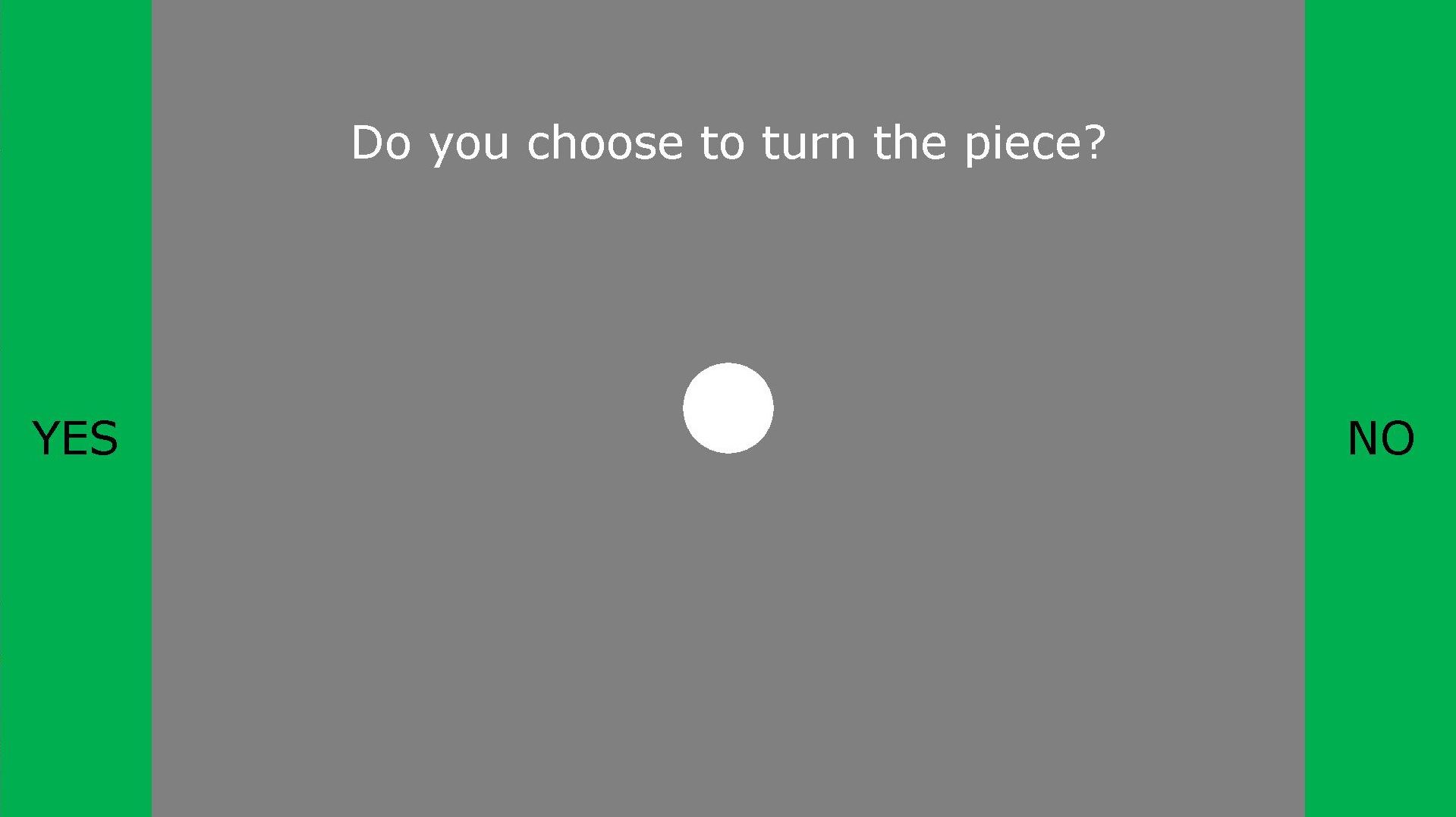}
\caption{\textbf{SSVEP-Based EEG Brain-Computer Interface used by the Senders and the Receiver.} Participants conveyed their decisions regarding whether or not to rotate the current block by controlling a cursor (white filled circle) using EEG based steady state visually evoked potentials (SSVEPs). Participants focused on a flashing LED to the left of the screen (depicted as a circle attached to the screen in  Figure~\ref{fig:brainnet}) to move the cursor leftwards towards the ``Yes'' side. Focusing on the LED to the right of the screen (flashing at a different frequency) caused the cursor to move rightwards towards the "No" side. If the cursor reached the green "Yes" bar, the interface interpreted the participant's decision to be rotation of the block (by 180 degrees). If the cursor reached the ``No'' bar, the interface took the participant's decision to be to keep the block’s current orientation.}
\label{fig:ssvep-screen}
\end{figure}

Figure~\ref{fig:brainnet} depicts the architecture of BrainNet. BrainNet relies on two well-known technologies: Electroencephalography (EEG)\cite{Nunez:2005} for non-invasively recording brain signals from the scalp and transcranial magnetic stimulation (TMS)\cite{OShea:2007} for non-invasively stimulating the visual cortex. The Senders convey their decisions of "rotate" or "do not rotate" by controlling a horizontally moving cursor (Figure~\ref{fig:ssvep-screen}) using steady-state visually-evoked potentials (SSVEPs) \cite{Vialatte:2010}: to convey a ``Rotate'' decision, Senders focused their attention on a ``Yes'' LED light flashing at 17 Hz placed on the left side of their computer screen; to convey a "Do Not Rotate" decision, they focused on the “No” LED light flashing at 15 Hz placed on the right side. These LEDs are depicted as circles attached to the screens in  Figure~\ref{fig:brainnet}. The cursor position provided real-time visual feedback to the Senders. The direction of movement of the cursor was determined by comparing the EEG power at 17 Hz versus 15 Hz, with a higher power at 17 Hz over that at 15 Hz moving the cursor towards the left side near the “Yes” LED, and vice-versa for the "No" LED. A ``Rotate'' (``Do Not Rotate'') decision was made when the cursor hit the side of the screen appropriately marked   "YES" ("NO") (see Figure~\ref{fig:ssvep-screen}). In trials where the cursor did not reach either side of the screen due to trial time elapsing, the decision closest to the last location of the cursor was chosen as the subject's decision.

The decisions of the two Senders were sent to the Receiver’s computer through a TCP/IP network and were further translated into two pulses of transcranial magentic stimulation (TMS) delivered sequentially to the occipital cortex of the Receiver. Each TMS pulse lasted 1 ms. An eight-second delay was enforced between the two  pulses  to remain within the strictest safety guidelines of TMS stimulation\cite{Rossi:2009}. The intensity of the stimulation was set above or below the threshold at which the Receiver could perceive a flash of light known as a phosphene: a ``Yes'' response was translated to an intensity above the threshold, and “No” was translated to an intensity below the threshold. During each round of trials, the Receiver always received the decision from one Sender first, then the other. The screen the Receiver saw also had visual prompts to remind them whose decision the current TMS stimulation was conveying.  Receivers made their decision based on whether a phosphene was perceived and conveyed their decision (rotate or do not rotate) to the  game using the same SSVEP-based procedure used by both Senders. After the game state was updated, the trial moved into the second round and the above process was repeated. At the end of each trial, all three subjects received feedback on the result of the trial (Figure~\ref{fig:two-step}, bottom row).

\subsection*{Differential Reliability of Senders}

When the decisions from the two Senders do not agree with each other, the Receiver must decide which Sender to trust. To investigate whether the Receiver can learn the reliability of each Sender and choose the more reliable Sender for making decisions,  we designed the system to deliberately make one of the Senders less accurate than the other. Specifically, for each session, one Sender was randomly chosen as the ``Bad'' Sender and, in 10 out of sixteen trials, this Sender's decision when sent to the Receiver was forced to be incorrect, both in the first and second round of each trial. 

\subsection*{EEG Procedure for Senders}

Each Sender performed the task in a dedicated room in front of a 21" LCD monitor, with two Arduino-controlled LED lights attached to the left and right outer frames of the monitor for eliciting SSVEPs. EEG signals were recorded through an 8-channel OpenBCI Cyton system (OpenBCI: Brooklyn, NY) with a sampling rate of 250Hz at a resolution of 16 bits. Signals were acquired from gold-plated electrodes and a layer of electro-conductive paste was applied between each electrode and the participant’s scalp. For the experimental session, three electrodes were set up along the midline in a custom montage with the signal recorded from one occipital electrode (location Oz in the 10–10 placement system) and two frontal electrodes (locations AFz and FCz in the 10–10 system) used as the ground and reference, respectively. 

Incoming EEG data was passed through a 4th-order Butterworth filter \cite{Sorrentino:2007} between 0 and 30 Hz to remove signal drifting and line noise. The time-series EEG data was then divided into 1-second epochs and transformed to the frequency domain using Welch's method\cite{Welch:1967}. The intention to rotate the falling block or not was decoded by comparing the power at 17 Hz and 15 Hz obtained from Welch's method. The final decision was made by tallying up the number of epochs in which the greatest power was recorded at either 17 Hz or 15 Hz over a 10-second period. Signal processing and data storage were managed through a custom software library developed by two of the authors (LJ and DL).

There is no prior training required for controlling the cursor using SSVEP. During the experiment, the Sender's monitor displays either the cursor-control interface or a gray background with a text prompt indicating that the Receiver is making a decision.

\subsection*{TMS Procedure for the Receiver}

Participants playing the role of the Receiver came in for two consecutive sessions. During the first session, as part of informed consent, they were asked to complete a TMS safety screening questionnaire, aimed at identifying potential conditions (such as family history of seizures or frequent migraines) that might represent potential risk factors for adverse side effects of TMS. 
No participant was rejected for failing the safety questionnaire. In addition to the safety screening, all Receivers underwent a procedure to determine their absolute phosphene threshold, that is, the minimum amount of stimulation necessary to elicit the perception of an induced phosphene 50\% of the time. The absolute threshold was assessed using the PEST method\cite{Taylor:1967}. The absolute threshold was then used as the starting point to identify the stimulation levels associated with the binary ``Rotate'' and ``Do Not Rotate'' decisions. Starting from the absolute threshold, the stimulation intensity was first adjusted upwards in increments of 5\% until phosphenes could be elicited for 10 consecutive pulses; this value was then used for conveying a ``Rotate'' decision from a Sender. Then, starting from the absolute threshold value, the stimulation intensity was lowered in 5\% increments until no phosphene was elicited for 10 consecutive pulses. This value was then used to convey a ``Do Not Rotate'' decision from a Sender. During both the testing session and the experimental session, TMS was delivered through a 70-mm Figure-8 Alpha coil (Magstim, UK) positioned over the left occipital lobe in a location corresponding to site O1 in the 10-20 system. The coil was positioned flushed to the head, with the handle parallel to the ground and extending towards the left. The coil was attached to a SuperRapid2 magnetic stimulator (Magstim, UK). 
The maximum intensity of the electric field for our TMS equipment is 530 V/m, and with our coil, the maximum intensity of the induced magnetic field is 2.0 T.

\subsection*{EEG Procedure for the Receiver}

The EEG procedure for the Receiver was identical to that used for the Senders, except that the signal was acquired from a BrainAmp system (BrainVision, Berlin, Germany) with a sampling rate of 5000 Hz and a resolution of 20 bits. The system was equipped with a DC voltage amplifier to reduce signal distortions due to the TMS pulses. Participants wore a standard 32-channel headcap, using AFz and FCz as the ground and reference, respectively. As in the case of the Senders, only data from the Oz channel was recorded. After downsampling to 500Hz, the incoming data underwent the same preprocessing steps described above for the Senders.

\bibliography{ref}

\section*{Acknowledgements}

This work is made possible by a W. M. Keck Foundation Award to AS, CP, and RPNR, and a Levinson Emerging Scholars Award to LJ. RPNR was also supported by NSF grant no.\ EEC-1028725 and a CJ and Elizabeth Hwang Endowed Professorship. We thank Nolan Strait for software testing. 

\section*{Author Contributions Statement}

RPNR, AS, CP conceived the experiment,  LJ, JA conducted the experiment, LJ, DL implemented the software, LJ and AS analyzed the results. All authors were involved in writing and reviewing the manuscript. 

\section*{Data Availability}

Experiment data and code are available upon request.

\section*{Competing Interests}

The authors declare no competing interests.

\section*{Corresponding Author}

Correspondence to Rajesh P. N. Rao (email: rao@cs.washington.edu).

\newpage

\end{document}